\documentclass[preprint]{ptephy_v1}

\preprintnumber{XXXX-XXXX} 
\usepackage{hyperref}

\begin{document}
\title{Phenomenological study of the charged particles production in pPb 
collisions at $\sqrt{s_{\rm{NN}}}$ = 5.02 TeV}

\author{Kapil Saraswat}
\affil{Institute of Physics, Academia Sinica, Taipei, Taiwan.
\email{kapilsaraswatbhu@gmail.com}}
\author{Deependra Singh Rawat}
\affil{School of Allied Sciences (Physics), Graphic Era Hill University,
Bhimtal Campus, Sattal Road, Nainital, India.
\email{dsrawatphysics@gmail.com}}
\author{Akash Pandey}
\affil{Department of Physics, Indian Institute of Technology, Mumbai, India.
\email{pandeyakash142@gmail.com}}
\author{Venktesh Singh}
\affil{Department of Physics, School of Physical Science, 
Central University of South Bihar, Gaya, India.
\email{venktesh@cusb.ac.in}}
\author{H. C. Chandola}
\affil{Department of Physics (UGC-Centre of Advanced Study),
Kumaun University, Nainital, India.
\email{chandolaharish@gmail.com}}

\begin{abstract}
We have studied transverse momentum ($p_{\rm{T}}$) spectra of charged hadrons in 
various pseudo-rapidity ranges for p-Pb collisions at
$\sqrt{s_{\rm{NN}}}$ = 5.02 TeV.
The medium effects such as collective flow and energy loss resulting from
heavy-ion collisions have also been investigated using modified Tsallis
distribution function over a wide range of $p_{\rm{T}}$ that indicates the 
transverse collective flow at low and intermediate $p_{\rm{T}}$ range and
in-medium energy loss in high $p_{\rm{T}}$ range.
\end{abstract}

\subjectindex{Charged Particles, Quark-Gluon Plasma, Collective Flow, Energy loss}
\maketitle

\section{Introduction}
The RHIC (Relativistic Heavy Ion Collider) \cite{phenix05, star05} and the LHC
(Large Hadron Collider) \cite{lhc08} are best experimental facilities that are
designed to investigate an exotic state of matter dubbed as Quark-Gluon Plasma
(QGP) \cite{shk80} which needs to be explore at both perturbative and
nonperturbative \cite{ks0, ks1, ks2, ks3, ymc1, hcc09, dsr18, hcc20, dsr21}
sector of QCD. The QGP is a color conducting state of matter in which quarks and
gluons are deconfined under extreme conditions similar to the matter in the early
universe around 1 $\mu$-sec. just after the big-bang \cite{olive91, schwarz03}.
There are numerous heavy-ion collision \cite{bh97} events like Au-Au, Cu-Cu, Cu-Au
etc in RHIC and p-Pb, Pb-Pb, Xe-Xe etc in LHC among them p-p collisions are taken
as baseline measurements. The particle production mechanism in p-p collisions are
charcterized by spectra of transverse momentum that provides insight into the state
of matter at freeze-out conditions. Recent studies of high multiplicity p-p
collisions at LHC \cite{cms10, alice17} predicts the formation of QGP like medium
in heavy-ion collisions.

In high energy heavy-ion collisions the momentum distribution of emitted hadrons
exhibit additional medium effects such as collective flow, recombination and jet
quenching etc. The collective flow \cite{hn04} of hadrons, caused by the expansion 
of a hot \cite{rf, rf1} dense matter formed in the head-on collision and
jet-quenching \cite{wang04}, where high energetic beams of particles loses enrgy 
on traversing through such dense medium. The hadron $p_{\rm{T}}$-spectra that
provides the particle production mechanism in p-p collisions are analyzed by
Tsallis distribution \cite{Tsallis:1987eu, Biro:2008hz} which is characterized 
by two parameters $T$ and $q$ \cite{adare11}. The parameter $T$ corresponds to
kinetic energy freez out temperature at which elastic collisions among particles
are stops to modify the spectral distribution of different hadrons whereas the
nonextensive parameter q governs the degree of non thermalization and measures
the thermal fluctuations in the system. The functional form of Tsallis
distribution which models system near thermal equilibrium, resembles with
Hagedorn's \cite{Hagedorn:1983wk} power law function which is employed in
hard scattering processes of QCD \cite{Hagedorn:1983wk, bb74}.

In the present paper we have investigated the $p_{\rm{T}}$-distribution of
charged hadrons in p-Pb collisions at $\sqrt{s_{\rm{NN}}}$ = 5.02 TeV measured
by the CMS experiment in the various pseudo-rapidity ranges. Using the modified
Tsallis distribution the transverse flow and in-medium energy loss of charged
particles has been described. The statistical and systematic errors are added
in quadrature and are incorporated in the fitting process. 

\section{Tsallis/Hagedorn distribution function and its modification}
In heavy-ion collisions, the transverse mass
($m_{\rm{T}} = \sqrt{p^{2}_{\rm{T}} + m^{2}}$) distribution of the produced particles
can be described by the Hagedron function which is a QCD-inspired summed power
law \cite{Hagedorn:1983wk} given as :
\cite{Hagedorn:1983wk} given as
\begin{eqnarray}
E~\frac{d^{3}N}{dp^{3}} = C~ \Bigg(1 + \frac{m_{\rm{T}}}{p_{0}}\Bigg)^{-n}~.
\label{Hag}
\end{eqnarray}

Eq.(\ref{Hag}) shows both the bulk spectra in the low $m_{\rm{T}}$ region and
the particles produced in QCD hard scatterings reflected in the high $p_{\rm{T}}$
region.
We compare this function (Eq.\ref{Hag})  with the Tsallis distribution
\cite{Tsallis:1987eu, Biro:2008hz} of thermodynamic origin which is given by
\begin{eqnarray}
E \frac{d^{3}N}{dp^{3}} = 
C_n ~ m_{\rm{T}} ~ \Bigg(1 + (q-1) \frac{m_{\rm{T}}}{T} \Bigg)^{-1/(q - 1)}~.
\label{Tsallis}
\end{eqnarray}

The Tsallis distribution [Eq.(\ref{Tsallis})] describes near-thermal systems in
terms of Tsallis parameter $T$ and the n0n-extensive parameter $q$ which measures
degree of non-thermalization \cite{Wilk:1999dr}.
The functions in Eq.~(\ref{Hag}) and in Eq.~(\ref{Tsallis}) have similar
mathematical forms with $n = 1/(q - 1)$ and $p_{0} = n~T$.
Larger values of $n$ correspond to smaller values of $q$.
Both $n$ and $q$ have been interchangeably used in Tsallis distribution
\cite{Biro:2008hz, adare11, Adare:2010fe, Cleymans:2012ya, Abelev:2006cs}.

Many theoretical and phenomenological study predict that  $n\sim4$
\cite{Blankenbecler:1975ct, Brodsky:2005fza} for quark-quark point scattering.
$n$ will be larger if multiple scattering centers are involved.
The study in Ref.~\cite{Zheng:2015mhz} suggests that both the forms given in
Eq.~(\ref{Hag}) and in Eq.~(\ref{Tsallis}) excellently describe the measured
data hadron spectra in pp collisions.
We use Eq.~(\ref{Tsallis}) in case of pp collisions. Tsallis/Hagedorn function
is able to describe $p_{\rm{T}}$ spectra in pp collisions at all RHIC and LHC
energies.

There have been many attempts to use the Tsallis distribution in heavy ion
collisions as well by including the transverse collective
flow ~\cite{Tang:2008ud, Khandai:2013fwa, Sett:2015lja} and in-medium energy
loss ~\cite{ks2}.
In the present work, we are using the modified Tsallis distribution ~\cite{ks2}
which includes the transverse flow in low $p_{\rm{T}}$ region and in-medium loss
at high $p_{\rm{T}}$ region.
The functional form of modified Tsallis distribution ~\cite{ks2} is given as : 

\begin{subequations} \label{modified_new_func_tsallis_distribution_function}
\begin{align} 
E \frac{d^{3}N}{dp^{3}} &= A_{1} \Bigg[\exp\left(-\frac{\beta  p_{\rm{T}}}{p_{1}}\right)
   + \frac{m_{\rm{T}}}{p_{1}}\Bigg]^{-n_{1}} ~ :~  p_{\rm{T}} < p_{\rm{T_{th}}} ~.
\label{new_func_tsallis_distribution_function} \\
E \frac{d^{3}N}{dp^{3}}  &= A_{2}~ \Bigg[\frac{B}{p_{2}}~\Bigg(\frac{p_{\rm{T}}}{q_{0}}\Bigg)^{\alpha}
                              + \frac{m_{\rm{T}}}{p_{2}}\Bigg]^{-n_{2}}~ :~  p_{\rm{T}} > p_{\rm{T_{th}}} ~. 
 \label{new_func_tsallis_distribution_function_second}
\end{align} 
\end{subequations}

Eq.~(\ref{new_func_tsallis_distribution_function}) shows the thermal and
collective behaviour of the hadton spectra with the parameters : the temperature
($T = p_{1}/n_{1}$) and the average transverse flow velocity
($\beta$)~\cite{Khandai:2013fwa}.\\
If we shift the transverse mass distribution in Eq.~(\ref{Hag}) by the energy 
loss $\Delta m_{\rm{T}}$ in the medium,
Eq.~(\ref{new_func_tsallis_distribution_function_second}) can be obatined. 
\begin{equation} 
E \frac{d^{3}N}{dp^{3}} =
A_2~ \Bigg[1 + \frac{m_{\rm{T}} +\Delta m_{\rm{T}}}{p_{2}}\Bigg]^{-n_{2}}~.
\label{tsallis_with_energyloss}
\end{equation}

The energy loss $\Delta m_{\rm{T}}$ is proportional to $p_{\rm{T}}$ at low $p_{\rm{T}}$.
The functional form of $\Delta m_{\rm{T}}$ has been taken from the
Ref.~\cite{Spousta:2016agr} which is given as 

\begin{equation}
\Delta m_{\rm{T}} = B ~ \Big(\frac{p_{\rm{T}}}{q_{0}}\Big)^\alpha~. 
\label{spousta_energy_loss}
\end{equation}

Here, the parameter $\alpha$ quantifies different energy loss regimes for light
quarks in the medium~\cite{Baier:2000mf, De:2011fe}.
The parameter $B$ is proportional to the medium size and $q_{0}$ is set 1GeV as
arbitrary scale.
Using Eq.~(\ref{spousta_energy_loss}) in Eq.~\ref{tsallis_with_energyloss} and
gnoring 1 we get Eq.~(\ref{new_func_tsallis_distribution_function_second})
applicable for high  $p_{\rm{T}}$. The parameter $p_{2}$ is not independent
parameter.
AT RHIC energy, the empirical energy loss in nuclear collisions is found to be
proportional to $p_{\rm{T}}$ \cite{Wang:2008se}. 

\section{Results and discussions}

\begin{figure}[htp]
\centering
\includegraphics[width=0.80\linewidth]{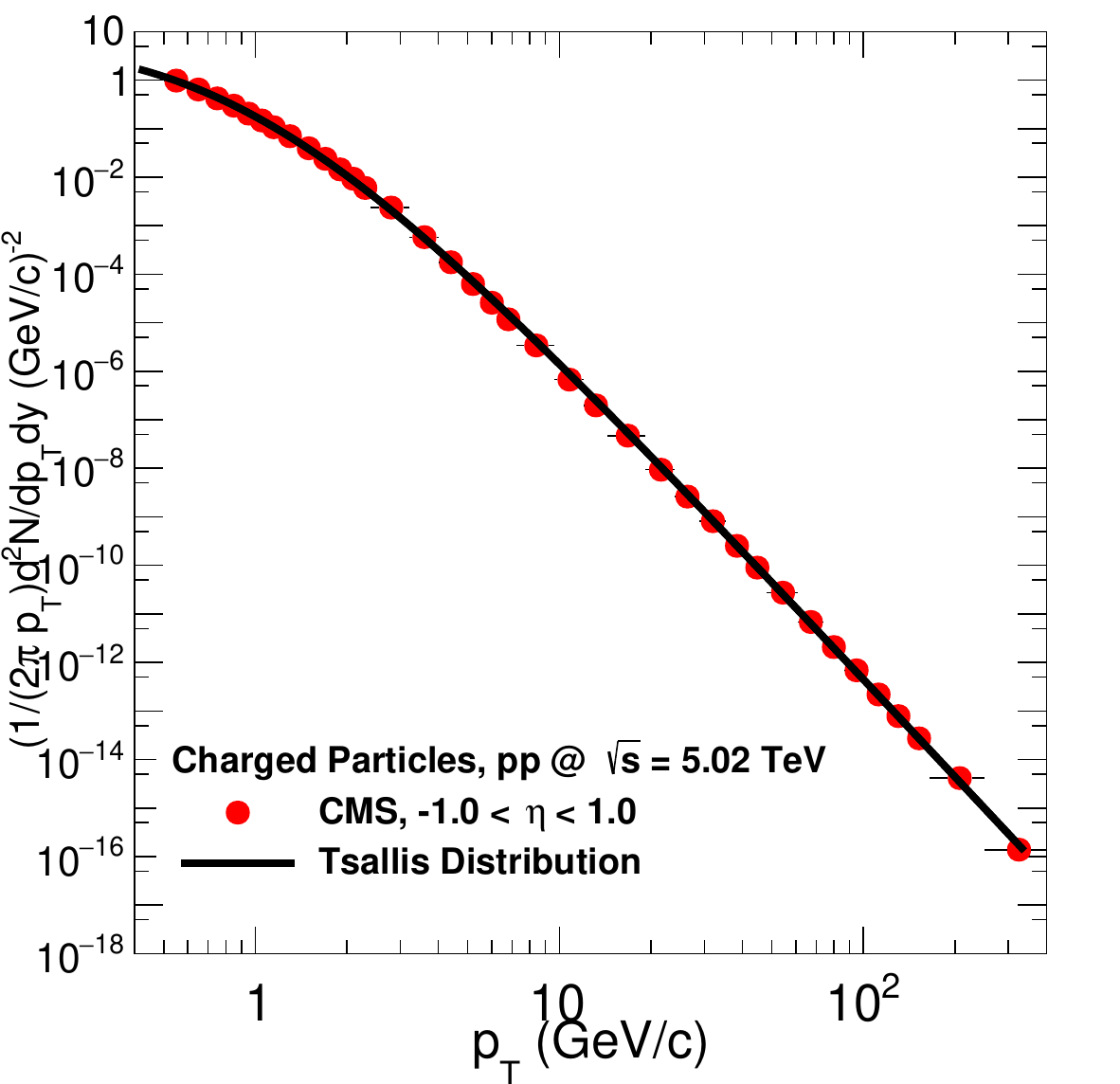}
\caption{The invariant yields of the charged particles as a function of 
transverse momentum $p_{\rm{T}}$ for pp collision at $\sqrt{s}$ = 5.02 TeV 
measured by the CMS experiment  \cite{CMS:2016xef}. The solid curves are
the fitted Tsallis distribution functions.}
\label{Figure1_pp_collision_cms}
\end{figure}

Figure (\ref{Figure1_pp_collision_cms}) shows the invariant yields of the charged
particles as a function of transverse momentum ($p_{\rm{T}}$) in psudo-rapidity
range $ - 1.0 < \eta < 1.0 $ for pp collisions at $\sqrt{s}$ = 5.02 TeV
measured by the CMS experiment \cite{CMS:2016xef}.
The solid curves are the Tsallis distributions fitted to the spectra.
The Tsallis distribution function gives good description of the data for both
the collision energies which can be inferred from the values  of
$\chi^{2}/\rm{NDF}$ given in the Table~(\ref{Table_one_pp_collision}).\\

\begin{figure}[htp]
\centering
\includegraphics[width=0.80\linewidth]{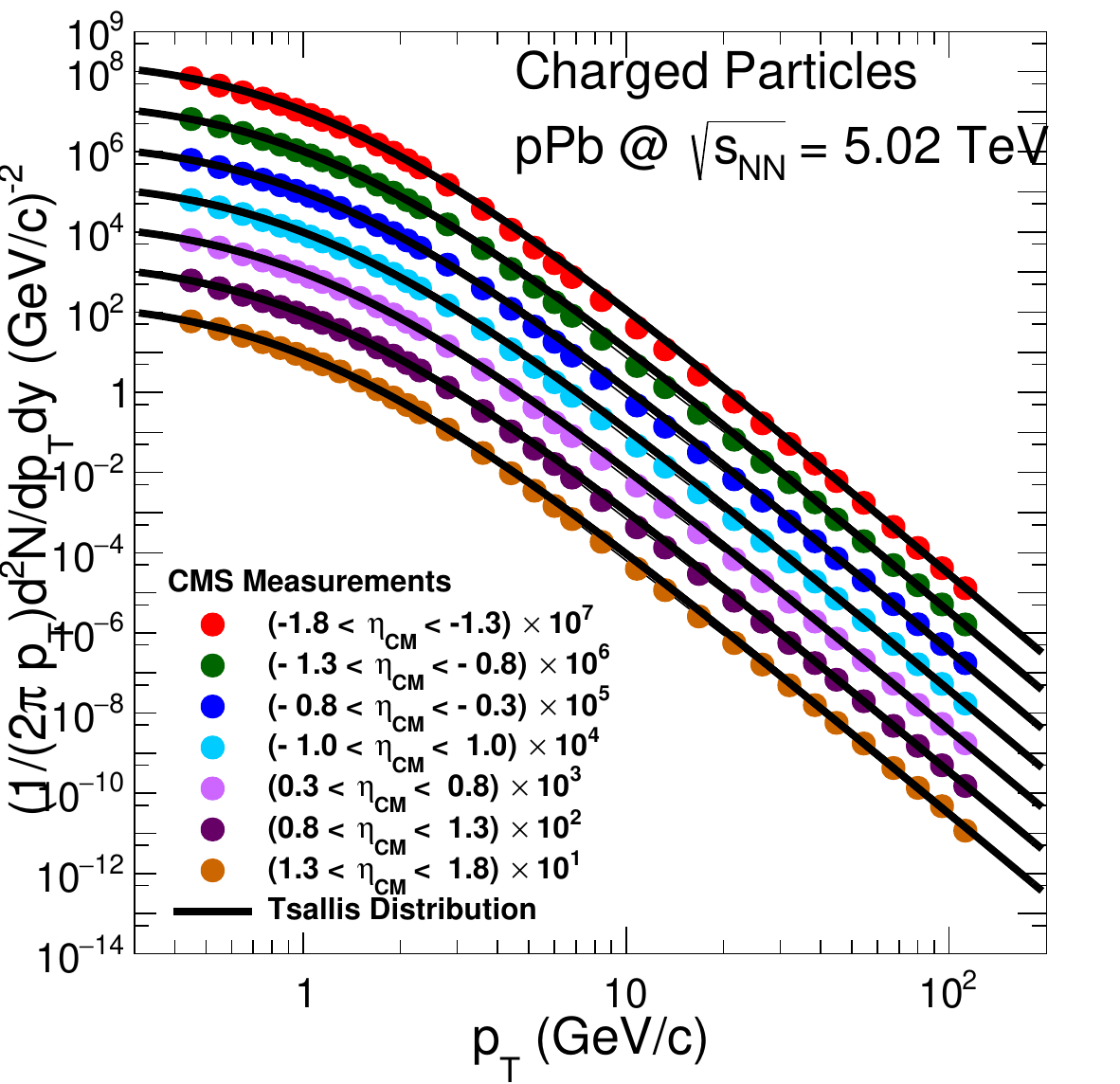}
\caption{The invariant yields of the charged particles  as a function of the 
transverse momentum $p_{\rm{T}}$ for pPb collisions at $\sqrt{s_{\rm{NN}}}$ =
5.02 TeV measured by the CMS experiment \cite{CMS:2015ved}. 
The solid curves are the fitted Tsallis distribution functions.}
\label{Figure2_pLead_502tev_tsallis}
\end{figure}

\begin{figure}[htp]
\centering
\includegraphics[width=0.80\linewidth]{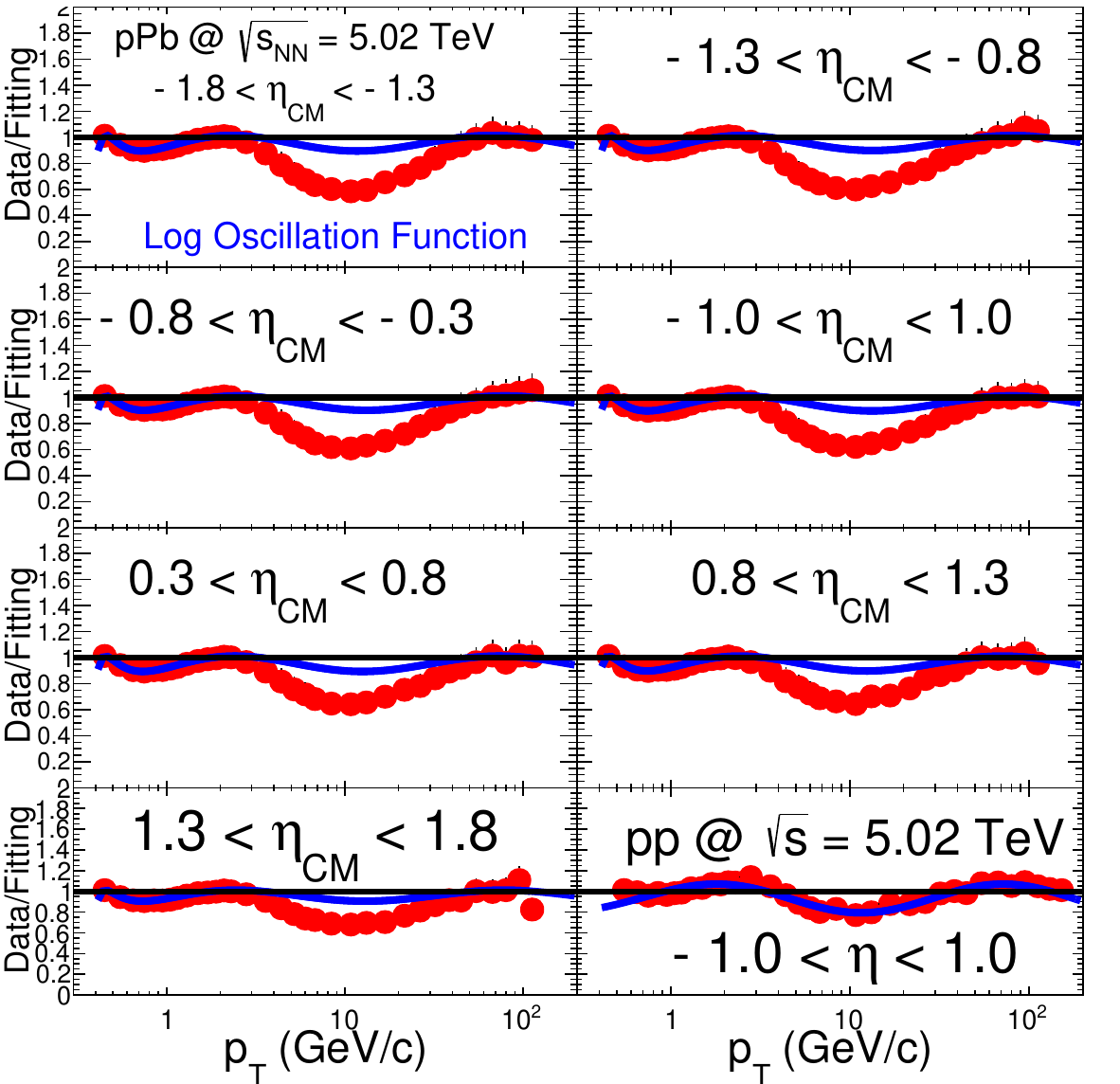}
\caption{The ratio of the charged particles yields data and their Tsallis fits
as a function of the transverse momentum $p_{\rm{T}}$ for pPb collisions at 
$\sqrt{s_{\rm{NN}}}$ = 5.02 TeV.
The solid curves are given by Eq.~\ref{log_oscillation_function_equation}.}
\label{Figure3_pLead_502tev_databyfit}
\end{figure}

Figure (\ref{Figure2_pLead_502tev_tsallis}) shows the invariant yields of the
charged particles as a function of $p_{\rm{T}}$  in various psudo-rapidity ranges
$(- 1.8 < \eta_{\rm{CM}} < - 1.3,~  - 1.3 < \eta_{\rm{CM}} < - 0.8,~ 
- 0.8 < \eta_{\rm{CM}} < - 0.3,~ - 1.0 < \eta_{\rm{CM}} < 1.0,~ 
0.3 < \eta_{\rm{CM}} < 0.8,~  0.8 < \eta_{\rm{CM}} < 1.3,~ 
0.8 < \eta_{\rm{CM}} < 1.3,~  1.3 < \eta_{\rm{CM}} < 1.8)$ 
for pPb collisions at $\sqrt{s_{\rm{NN}}}$ = 5.02 TeV
measured by the CMS experiment \cite{CMS:2015ved}.
The solid curves are the fitted Tsallis distributions.
Figure (\ref{Figure3_pLead_502tev_databyfit}) shows the ratio of the data
and the fitted Tsallis distribution as a function of $p_{\rm{T}}$ for pPb
collisions at $\sqrt{s_{\rm{NN}}}$ = 5.02 TeV. CMS measured data of pPb and
pp collisions show deviations from the fit which can be inferred from the
values  of $\chi^{2}/\rm{NDF}$ given in the
Table~(\ref{Table_two_pLead_collision_tsallis}).
The ratio of the data to the Tsallis fit shows a log oscillation function
in Ref.~\cite{Wilk:2014bia, Rybczynski:2014ura} and it can be parametrized
by a function of the form
\begin{eqnarray}
f(p_{\rm{T}}) = a + b~ \cos\big[c~\log(p_{\rm{T}} + d) + e\Big]~. 
\label{log_oscillation_function_equation}
\end{eqnarray}
Here $a$, $b$, $c$, $d$ and $e$ are the fit parameters.
From Figure (\ref{Figure3_pLead_502tev_databyfit}), we observe that the log
oscillation function with 5 parameters does not describe the deviation pattern.

\begin{figure}[htp]
\centering
\includegraphics[width=0.80\linewidth]{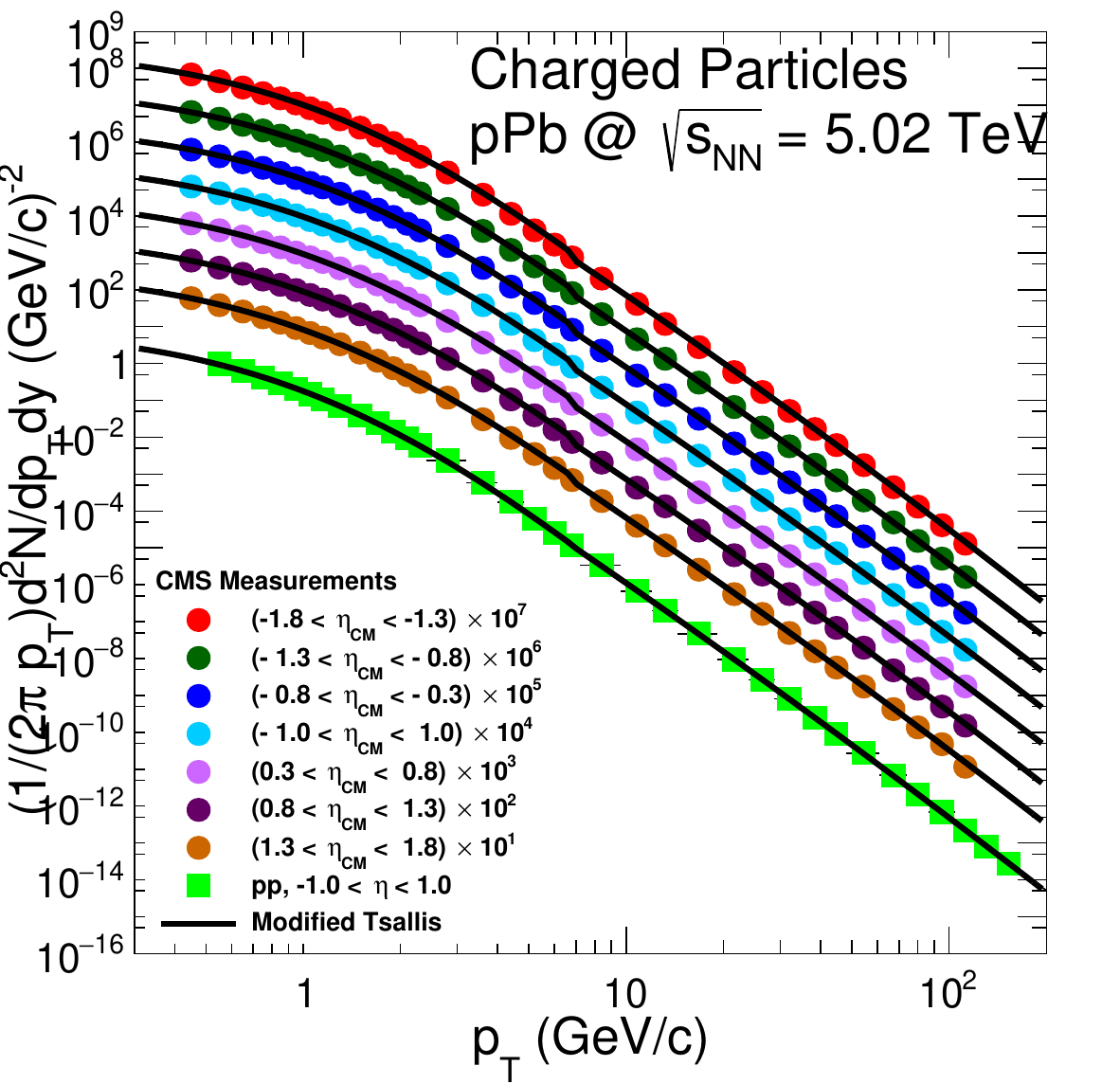}
\caption{The invariant yields of the charged particles  as a function of the 
transverse momentum $p_{\rm{T}}$ for pPb collisions at $\sqrt{s_{\rm{NN}}}$ =
5.02 TeV measured by the CMS experiment \cite{CMS:2015ved}.
The solid curves are the modified Tsallis distributions
(Eq. \ref{new_func_tsallis_distribution_function}).}
\label{Figure4_pLead_502tev_tsallis_modified}
\end{figure}

\begin{figure}[htp]
\centering
\includegraphics[width=0.80\linewidth]{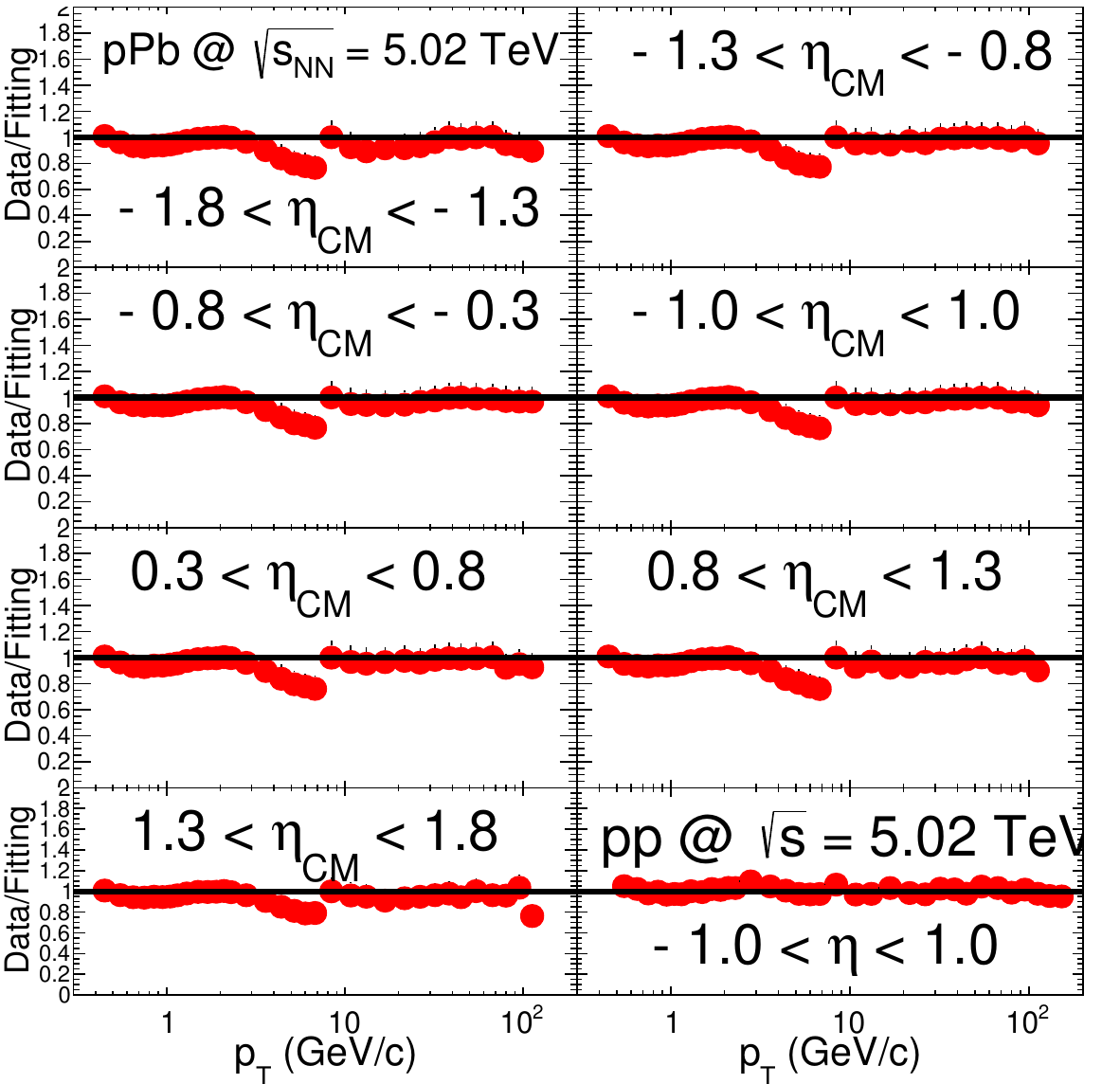}
\caption{The ratio of the charged particle yield data and the fit function  
(Modified Tsallis distribution Eq.~\ref{new_func_tsallis_distribution_function}) 
as a function of the transverse momentum $p_{\rm{T}}$ for pPb collisions at 
$\sqrt{s_{\rm{NN}}}$ = 5.02 TeV.}
\label{Figure5_pLead_502tev_databyfit_modified}
\end{figure}

Figure (\ref{Figure4_pLead_502tev_tsallis_modified}) shows the invariant yields
of the charged particles as a function of $p_{\rm{T}}$ in various psudo-rapidity 
ranges $(- 1.8 < \eta_{\rm{CM}} < - 1.3,~  - 1.3 < \eta_{\rm{CM}} < - 0.8,~ 
- 0.8 < \eta_{\rm{CM}} < - 0.3,~ - 1.0 < \eta_{\rm{CM}} < 1.0,~ 
0.3 < \eta_{\rm{CM}} < 0.8,~  0.8 < \eta_{\rm{CM}} < 1.3,~ 
0.8 < \eta_{\rm{CM}} < 1.3,~  1.3 < \eta_{\rm{CM}} < 1.8)$ for pPb collisions 
at $\sqrt{s_{\rm{NN}}}$ = 5.02 TeV measured by the CMS experiment \cite{CMS:2015ved}.
The solid curves are the modified Tsallis distributions given by
Eq.~(\ref{new_func_tsallis_distribution_function} and 
\ref{new_func_tsallis_distribution_function_second}). 
Figure (\ref{Figure5_pLead_502tev_databyfit_modified}) shows the ratio
of the data and the fit function by the modified Tsallis distribution
as a function of $p_{\rm{T}}$ for pPb and pp collisions at
$\sqrt{s_{\rm{NN}}}$ = 5.02 TeV. The ratio of the data and the fit function
shows that modified Tsallis distribution function gives excellent
description of the measured data in full $p_{\rm{T}}$ range for all the systems.
The parameters of the modified Tsallis distribution are given in the
Table~(\ref{Table_three_pLead_collision_tsallis_modified}).
The values of the first set of parameters ($n_{1}$, $p_{1}$, $\beta$)
are constant for differnt ramge of pseudo-rapidities.
While fitting the second function, we fix the parameter $n_{2} =7.71$
guided by pp value.
The exponent $\alpha$ which decides the variation of the energy 
loss of partons as a function of their energy remains same (=0.6).
In conclusion, the function given in Eq.~(
\ref{new_func_tsallis_distribution_function} and
\ref{new_func_tsallis_distribution_function_second}) gives excellent
description of the hadron spectra over wide range of $p_{\rm{T}}$
with its parameters indicating different physics effects in the
pPb collisions.

\clearpage
\section{Conclusion}
In the article, we analyze the CMS measured data of the transverse momentum
($p_{\rm{T}}$) spectra of the charged particles in pp collision at $\sqrt{s}$ =
5.02 TeV using Tsallis distribution function. Tsallis distribution function
describes the $p_{\rm{T}}$ spectra of the charged particles. 
We also analyze the $p_{\rm{T}}$ spectra of the charged particles for differnet
pseudo-rapidity range in pPb collisions at $\sqrt{s_{\rm{NN}}}$ = 5.02 TeV.
After analyzing the measured data, we found that $p_{\rm{T}}$ spectra in pPb
collisions show deviations from the Tsallis distribution function.
To solve this problem, we used the modified Tsallis distribution function which
incorporates the transverse flow in the low $p_{\rm{T}}$ region and in-medium
energy loss in the high $p_{\rm{T}}$ region.
The modified Tsallis distribution function excellently describe the $p_{\rm{T}}$ 
spectra in pp and pPb colliisons with its parameters which are quantify the
in-medium effects in both above systems.
In the low $p_{\rm{T}}$ region, the parameters $n_{1}, p_{1}$ and $\beta$ are
remain same for different pseudo-rapidity ranges.
We found that in the high $p_{\rm{T}}$ region, the parameter $\alpha$ which
describes the energy loss of the partons as a function of their remains within
the range 0.60 to 0.65. 

\section{Appendix}
\begin{table}[ht]
\caption{The parameters of the Tsallis function obtained by fitting the 
charged particles transverse momentum ($p_{\rm{T}}$) spectrum in pp collision 
at $\sqrt{s}$ = 5.02 TeV.}
\label{Table_one_pp_collision}
\begin{center}
\begin{tabular}{| c || c | c | c | c | c |} \hline
  Rapidity   &  $n$ & $q$ &   $T$  & $\frac{\chi^{2}}{\rm{NDF}}$ \\ 
  ($\eta$)   &      &     & (MeV)  &               \\ \hline \hline
  - 1.0 $<~ \eta ~<$ 1.0   & 7.71 $\pm$ 0.10  & 1.13    &   94.11 $\pm$ 7.84  & 0.05 \\ \hline 
\end{tabular}
\end{center}
\end{table}

\begin{table}[ht]
\caption{The parameters of the Tsallis function obtained by fitting the 
charged particles transverse momentum ($p_{\rm{T}}$) spectrum in pPb 
collisions $\sqrt{s_{\rm{NN}}}$ = 5.02 TeV.}
\label{Table_two_pLead_collision_tsallis}
\begin{center}
\begin{tabular}{| c || c | c | c | c | c |} \hline
  Rapidity   &  $n$ & $q$ &   $T$  & $\frac{\chi^{2}}{\rm{NDF}}$ \\ 
  ($\eta$)   &      &     & (MeV)  &               \\ \hline \hline

- 1.8 $<~ \eta_{\rm{CM}} ~<$ - 1.3   & 7.85 $\pm$ 0.00  & 1.13    &   112.79 $\pm$ 0.24  & 9.55 \\ \hline 

 - 1.3 $<~ \eta_{\rm{CM}} ~<$ - 0.8   & 7.82 $\pm$ 0.00  & 1.13    &   114.45 $\pm$ 0.24  & 9.09 \\ \hline 

 - 0.8 $<~ \eta_{\rm{CM}} ~<$ - 0.3   & 7.80 $\pm$ 0.00  & 1.13    &   115.31 $\pm$ 0.25  & 9.09 \\ \hline 

 - 1.0 $<~ \eta_{\rm{CM}} ~<$ 1.0   & 7.77 $\pm$ 0.00  & 1.13    &   113.35 $\pm$ 0.24  & 8.19 \\ \hline 

  0.3 $<~ \eta_{\rm{CM}} ~<$ 0.8   & 7.74 $\pm$ 0.00  & 1.13    &   111.68 $\pm$ 0.25  & 7.73 \\ \hline 

  0.8 $<~ \eta_{\rm{CM}} ~<$ 1.3   & 7.74 $\pm$ 0.00  & 1.13    &   109.62 $\pm$ 0.28  & 7.14 \\ \hline 

  1.3 $<~ \eta_{\rm{CM}} ~<$ 1.8   & 7.73 $\pm$ 0.00  & 1.13    &   106.92 $\pm$ 0.24  & 6.36 \\ \hline 

\end{tabular}
\end{center}
\end{table}

\begin{table}[ht]
 \caption{The parameters of the modified Tsallis function
 Eq.~\ref{new_func_tsallis_distribution_function} obtained by 
fitting the charged particle spectra in pPb collisions at $\sqrt{s_{NN}}$ = 5.02 TeV.}
\label{Table_three_pLead_collision_tsallis_modified}
\begin{center}
\resizebox{1.0\textwidth}{!}{
\begin{tabular}{|c || c | c | c | c | c | c | c  |}  \hline
 System   &  Rapidity   & $n_{1}$       & $p_{1}$         & $\beta$         &  $\alpha$      & $B$       & $\frac{\chi^{2}}{\rm{NDF}}$  \\  
          & ($\eta$)  &              & (GeV/$c$)       &                 &                & (GeV/$c$)      & \\ \hline \hline
 pPb    &   - 1.8 $<~ \eta_{\rm{CM}} ~<$ - 1.3  & 7.23 $\pm$ 0.30 & 1.16 $\pm$ 0.13 & 0.14 $\pm$ 0.05 & 0.60 $\pm$ 0.00 & 3.10 $\pm$ 0.02 & 2.91 \\ \hline 

   pPb    &   - 1.3 $<~ \eta_{\rm{CM}} ~<$ - 0.8  & 7.30 $\pm$ 0.03 & 1.22 $\pm$ 0.01 & 0.12 $\pm$ 0.01 & 0.64 $\pm$ 0.00 & 3.50 $\pm$ 0.03 & 2.62 \\ \hline 

   pPb    &   - 0.8 $<~ \eta_{\rm{CM}} ~<$ - 0.3  & 7.04 $\pm$ 0.31 & 1.12 $\pm$ 0.12 & 0.16 $\pm$ 0.05 & 0.62 $\pm$ 0.00 & 3.43 $\pm$ 0.03 & 2.43 \\ \hline 

   pPb    &   - 1.0 $<~ \eta_{\rm{CM}} ~<$ 1.0  & 7.04 $\pm$ 0.27 & 1.12 $\pm$ 0.11 & 0.14 $\pm$ 0.05 & 0.63 $\pm$ 0.00 & 3.53 $\pm$ 0.02 & 2.63 \\ \hline 

   pPb    &   0.3 $<~ \eta_{\rm{CM}} ~<$ 0.8  & 6.89 $\pm$ 0.33 & 1.06 $\pm$ 0.13 & 0.15 $\pm$ 0.06 & 0.64 $\pm$ 0.00 & 3.74 $\pm$ 0.04 & 2.66 \\ \hline 

   pPb    &   0.8 $<~ \eta_{\rm{CM}} ~<$ 1.3  & 6.90 $\pm$ 0.26 & 1.05 $\pm$ 0.11 & 0.14 $\pm$ 0.05 & 0.60 $\pm$ 0.00 & 3.33 $\pm$ 0.03 & 2.67 \\ \hline 

   pPb    &   1.3 $<~ \eta_{\rm{CM}} ~<$ 1.8  & 6.85 $\pm$ 0.27 & 1.01 $\pm$ 0.11 & 0.13 $\pm$ 0.05 & 0.63 $\pm$ 0.00 & 3.51 $\pm$ 0.03 & 2.34 \\ \hline 

   pp    &  -  1.0 $<~ \eta ~<$ 1.0  & 7.14 $\pm$ 2.59 & 0.99 $\pm$ 0.69 & 0.14 $\pm$ 0.10 & 0.65 $\pm$ 0.12 & 3.32 $\pm$ 1.06 & 0.01 \\ \hline 
\end{tabular}}
\end{center}
\end{table}

\clearpage 


\end{document}